\documentclass[pra,aps,12pt,floatfix,notitlepage,nofootinbib,superscriptaddress,longbibliography]{revtex4-2}

\usepackage{graphicx,bm,color}
\usepackage{array}
\usepackage{amsmath,amsfonts,amssymb}
\usepackage[utf8]{inputenc}
\usepackage[T1]{fontenc}
\usepackage{multirow}

\usepackage{tikz}
\usetikzlibrary{calc,patterns,decorations.pathmorphing,decorations.markings}

\usetikzlibrary{tikzmark}
\usepackage{chngcntr}
\counterwithin*{paragraph}{subsubsection}

\newcommand{\hk}[1]{{#1}}
\usepackage{geometry}
\usepackage[breaklinks=true]{hyperref}
\hypersetup{
    colorlinks,
    linkcolor={red!50!black},
    citecolor={blue!90},
    urlcolor={blue!80!black}
}

\newcommand{\be}{\begin{equation}}
\newcommand{\ee}{\end{equation}}
\newcommand{\bea}{\begin{eqnarray}}
\newcommand{\eea}{\end{eqnarray}}

\newcommand{\meanv}[1]{\langle #1 \rangle}

\newcommand{\bb}[1]{\left( #1 \right)}

\newcommand{\bbcro}[1]{\left[ #1 \right]}
\newcommand{\bbcror}[1]{\left. #1 \right]}

\newcommand{\bbaco}[1]{\left\{ #1 \right\}}

\newcommand{\ii}{\textrm{i}}
\newcommand{\dd}{\textrm{d}}
\newcommand{\eee}{\textrm{e}}
\newcommand{\rr}{\textbf{r}}

\newcommand{\qq}{\textbf{q}}

\newcommand{\KK}{\textbf{K}}

\newcommand{\yy}{\textbf{y}}
\newcommand{\pp}{\textbf{p}}

\newcommand{\zero}{\textbf{0}}

\newcommand{\ket}[1]{|#1\rangle}
\newcommand{\bra}[1]{\langle #1|}

\newcommand{\kF}{k_{\rm F}}
\newcommand{\eF}{\epsilon_{\rm F}}
\newcommand{\TF}{T_{\rm F}}
\newcommand{\tF}{t_{\rm F}}
\newcommand{\pF}{p_{\rm F}}
\newcommand{\vF}{v_{\rm F}}
\newcommand{\EF}{\epsilon_{\rm F}}

\newcommand{\upa}{\uparrow}
\newcommand{\dwa}{\downarrow}

\newcommand{\eqqref}[1]{Eq.~\eqref{#1}}
\newcommand{\eqqrefs}[2]{Eqs.~\eqref{#1}--\eqref{#2}}
\newcommand{\eqqrefss}[2]{Eqs.~(\ref{#1}--\ref{#2})}

\usepackage{tikz}
\usetikzlibrary{shapes.geometric, positioning, shadings, arrows.meta}
\usetikzlibrary{calc,patterns,decorations.pathmorphing,decorations.markings}

\tikzset{->-/.style={decoration={
  markings,
  mark=at position .5 with {\arrow{>}}},postaction={decorate}}}
\tikzset{-<-/.style={decoration={
  markings,
  mark=at position .5 with {\arrow{<}}},postaction={decorate}}}
\tikzset{phantom->-/.style={decoration={
  markings,
  mark=at position .5 with {\arrow[scale=2]{>}}},postaction={decorate}}}

\tikzset
  {
    ,my bubble/.style = 
      {
        ,draw=#1!70
        ,fill=#1!10
        ,ellipse
        ,inner sep=1pt
        ,minimum width=2em
        ,minimum height=2em
        ,align=center
      }
    ,my end/.style =
      {
        ,draw=#1!70
        ,top color=#1!10
        ,bottom color=#1!50
        ,minimum height=6em
        ,text width=6em
        ,inner sep=0pt
        ,align=center
      }
    ,my arrow/.style =
      {
        ,>=Stealth
        ,->
        ,draw=black
      }
  }
  
\tikzset{serpent/.style={decoration={snake},postaction={decorate}}}

\begin{document}
\title{The crossover from classical to quantum transport in a weakly-interacting Fermi gas}
\author{Hadrien Kurkjian}
\affiliation{Laboratoire de Physique Théorique de la Matière Condensée, Sorbonne Université, CNRS, 75005, Paris, France}
\email{hadrien.kurkjian@cnrs.fr}

\begin{abstract}
We present an exact solution of the quantum kinetic equation of a weakly-interacting two-component unpolarized Fermi gas in the crossover from the degenerate Fermi-liquid regime to the classical Boltzmann gas. We construct families of orthogonal polynomials tailored to each angular momentum channel, enabling a fast and systematically improvable decomposition of the phase-space distribution. This approach yields accurate, non-variational predictions for the shear viscosity, thermal diffusivity, and spin diffusivity to leading order in the scattering length. We demonstrate that the commonly used relaxation-time approximation fails at low temperature—by up to 25\%. Our method provides a numerically efficient framework for benchmarking transport in strongly correlated regimes and for simulating the kinetics of quantum gases beyond hydrodynamics.
\end{abstract}

\maketitle

\section*{Introduction}

A weakly-interacting Fermi gas smoothly interpolates
between a Fermi liquid at low temperatures,
and a Boltzmann gas at high temperatures. 
All along this transition from a quantum degenerate
to a classical gas, the dynamics can be captured by a kinetic
equation; as illustrated in Fig.~\ref{schema}, this contrasts with the strongly-interacting
regime where the Fermi liquid and the high-temperature gas are separated
by a strongly-correlated regime in which the Bogoliubov–Born–Green–Kirkwood–Yvon
(BBGKY) hierarchy cannot be truncated \cite{Bonitz1998,Kira2015,eindhoven}.

Exploiting the privileged situation of the weakly-interacting gas, one can work out accurate 
predictions of the transport properties, that can be used, either experimentally, theoretically 
or numerically, as benchmarks for the strongly-correlated regime. 
\hk{In a weakly-interacting gas, where the collision probability
coincides with the bare interparticle interaction, the only
remaining difficulty is to solve the kinetic equation without approximations.
For the transport coefficients, which characterize dissipation
in the hydrodynamic limit, this means solving a set of linear integral equations,
one for each transport coefficient.}

\hk{Several relaxation-time approximations (RTA) have been formulated to 
estimate the solution of these integral equations
without going through the inversion of the collision kernel.
Their common strategy is to replace the collision kernel (restricted to 
the subspace orthogonal to conserved quantities) by a simpler, typically 
diagonal, kernel.

The simplest formulation of the RTA is to replace the collision kernel by $1/\tau$ times the identity,
the so-called Bhatnagar–Gross–Krook (BGK) kernel; the collision time $\tau$ introduced in the BGK model is a phenomenological parameter,
tuned either to the mean collision time (the collision time averaged over the velocity distribution of the particles) or 
to the collision time at a characteristic energy (e.g. the Fermi energy \cite{Khalatnikov1958}). 
A better RTA, free from a phenomenological parameter, is obtained by adapting the diagonal approximation
to each transport channel using the projection of the collision kernel $\mathcal{M}$ onto the characteristic function 
$\ket{X}$ (normalized to unity) of the channel; in mathematical terms, this amounts to approximating $\bra{X}\mathcal{M}^{-1}\ket{X}$ by $1/\bra{X}\mathcal{M}\ket{X}$. 
This approximation is often called the (first) variational RTA, as it provides a lower bound
of the exact transport coefficients. It can be systematically improved into ``second'', ``third'' and so-on approximations \cite{ChapmanCowling} by using
a larger basis of orthogonal functions.}

\hk{Although they are not controlled by a small parameter, the RTAs provide qualitatively correct scalings of the transport coefficients in the high and
low temperature regimes. For contact or hard-sphere interactions, the variational RTA is also spectacularly successful
in the classical, high-temperature regime: the first approximation is only
a few percents away from the exact solution, and the convergence of the higher order 
approximations is extremely fast \cite{SmithJensen,ChapmanCowling}. 
However, it is known from \textit{e.g.} the Fermi liquid
literature that the RTAs become inaccurate at low temperatures \cite{Sykes1968,Wilkins1968},
and that the convergence with the number of trial functions \cite{disphydro} drastically slows down.
While the errors of the RTA in the classical-to-quantum crossover 
were carefully quantified for Bose gases \cite{Reichl2013}, studies on Fermi gases
were primarily concerned with the high-$T$ limit at unitarity \cite{Schafer2010,Smith2005,Smith2005b}. 
Deviations from the RTA below $\TF$ at weak coupling were computed by Ref.~\cite{Wu2019}, but the discussion
was unfortunately limited to $T>0.2\TF$, at the threshold of a rapid increase of those deviations.}

\hk{Here, we argue that RTAs can easily be discarded to solve the kinetic
equation of a weakly-interacting Fermi gas at any temperature. This is related
to the existence of a natural basis of orthogonal functions
which enables a fast and efficient numerical inversion of the collision kernel.}
We apply this method to obtain the shear viscosity $\eta$, thermal diffusivity $\kappa$ and spin diffusivity
$D$ of an \hk{unpolarized two-component} Fermi gas with contact interactions as a function of $T/\TF$, 
and to leading order in the s-wave scattering length $a$.
We observe a sharp increase of the error of the variational RTA from a few percents in the classical regime
\cite{SmithJensen}, to 10\%-30\% in the quantum regime \cite{Sykes1968,Wilkins1968,disphydro,yaleexp}.
\hk{Our results apply to temperatures above the superfluid critical temperature $T_c$; on the BCS
side ($a<0$), the critical temperature is exponentially reduced with $\kF a$, which opens a wide Fermi liquid regime
at $T_c< T\ll \TF$; on the BEC side ($a>0$), the superfluid phase can be avoided
as long as the  atoms do not form bosonic dimers and the gas remains on the metastable branch \cite{Castin2004} corresponding to a weakly-repulsive
Fermi liquid.}

The exact solution of the linearized Boltzmann equation is obtained by expanding 
the momentum distribution $n(\pp)$ on several bases of orthogonal polynomials. Restricting ourselves
to distributions rotationally invariant about the excitation wavevector $\qq$, we use Legendre polynomials to describe the angular
dependence of $n$; for the radial dependence, we construct families of polynomials $\{Q_n^l(p)\}_{n\in\mathbb{N}}$
adapted to both the temperature and to the partial wave $l$. \hk{This generalizes to finite nonzero temperatures 
the basis of Sonine (or generalized Laguerre) polynomials of the Boltzmann gas \cite{ChapmanCowling,Schafer2010}, and the 
orthogonal polynomials specific to the Fermi liquid regime \cite{Gran2023,devvisco}.}
This very efficient decomposition of the particle distribution could
be used for a fast and accurate numerical solution of the kinetic equation,
to study \textit{e.g.}, time-dependent evolutions or nonlinear effects.
The Fortran 90 source code of the numerical calculation of $\eta(T/\TF)$, $\kappa(T/\TF)$
and $D(T/\TF)$ is available in the online repository \cite{vtemp}.

\section{Transport equation of a weakly-interacting Fermi gas}

\subsection{Weakly-interacting Fermi gas}

We consider a gas of spin-$1/2$ fermions of mass $m$ and chemical 
potential $\mu$ trapped in a cubic volume $L^3$ and interacting via 
contact interactions, characterized by the coupling constant $g$. 
The total equilibrium density \hk{$\rho_{\rm eq}=\rho_{\rm eq,\upa}+\rho_{\rm eq,\dwa}$} fixes the Fermi units
\be
\pF=(3\pi^2\rho_{\rm eq})^{1/3},\qquad \EF=\frac{1}{\tF}=\frac{\pF^2}{2m}
\ee
(we use $\hbar=k_{\rm B}=1$ throughout). 
The Hamiltonian of the system reads:
\begin{eqnarray}
\hat{H}&=&\hat{H}_0+\hat{V}, \\ \hat{H}_0&=&\sum_{\pp\in\mathcal{D},\sigma} 
{\frac{p^2}{2m}} \hat a_{\pp\sigma}^\dagger \hat a_{\pp\sigma} \\
\hat{V}&=&g\int \dd^3r \hat\psi_\upa^\dagger(\rr)  \hat\psi_\dwa^\dagger(\rr) \hat\psi_\dwa(\rr) \hat\psi_\upa(\rr)=\frac{g}{L^3}\sum_{\pp_1,\pp_2,\pp_3,\pp_4\in
\mathcal{D}} \delta_{\pp_1+\pp_2}^{\pp_3+\pp_4} 
\hat a_{\pp_1\upa}^\dagger \hat a_{\pp_2\dwa}^\dagger \hat a_{\pp_3\dwa} \hat a_{\pp_4\upa}
\end{eqnarray}
where $\hat\psi_\sigma(\rr)$ is the field operator of fermions of spin $\sigma$, and $\hat a_{\pp\sigma}=\int \frac{\dd^3 r}{\sqrt{L^3}} \eee^{-\ii\pp\cdot\rr}\hat\psi_\sigma(\rr)$ its Fourier transform. The set of momenta $\mathcal{D}=(2\pi \mathbb{Z}/L)^3$ tends to
a continuum in the thermodynamic limit.
To first order in perturbation theory, we relate the 
coupling constant $g$ to the s-wave scattering length in the Born approximation \cite{Varenna,yaleexp}
\be
g=\frac{4\pi a}{m}
\ee

\subsection{Collisional dynamics of the particle distribution}
Unlike in strongly-interacting systems, the validity of the transport equation in a weakly-interacting gas 
is guaranteed at all temperatures and momenta by the smallness of the interactions. \hk{In a homogeneous state, one may
identify the phase-space distribution function with the momentum distribution of the {particles}
\be
\delta n_{\sigma}(\pp,t)=\meanv{\hat a^\dagger_{\pp\sigma} \hat a_{\pp\sigma}}-\meanv{\hat a^\dagger_{\pp\sigma} \hat a_{\pp\sigma}}_{\hat\varrho_{\rm th}}
\label{nhomogene}
\ee
The average value $\meanv{\hat O}=\text{Tr}(\hat\varrho\hat O)$ is taken in the out-of-equilibrium state of the system $\hat\varrho$,
and we have subtracted the thermal average $\meanv{\hat O}_{\hat\varrho_{\rm th}}$ in the equilibrium state
\be
\hat\varrho_{\rm th}(T)=\frac{\eee^{-\hat H/T}}{Z}
\ee
where $Z$ is the partition function. The equilibrium state populates the momentum modes
according to the Fermi-Dirac distribution:
\be
\meanv{\hat a^\dagger_{\pp\sigma} \hat a_{\pp\sigma}}_{\hat\varrho_{\rm th}}=n_{\rm eq}(\epsilon_{\sigma}(\pp))+O(g^2),\qquad n_{\rm eq}(\epsilon)=\frac{1}{1+\eee^{(\epsilon-\mu)/T}}
\ee 
where the particle dispersion $\epsilon_{\sigma}(\pp)$ is defined below (\eqqref{epsilonsigma}).
Within linear response theory, we expand the state $\hat\varrho$ of the system about thermal equilibrium 
\be
\hat\varrho=\hat\varrho_{\rm th}(T)+\delta\hat\varrho
\ee 
The fluctuation $\delta\hat\varrho$ is triggered by an external drive and is proportional to its intensity $U$.

In an inhomogeneous state, the momentum distribution \eqqref{nhomogene} becomes a phase-space distribution $n_{\sigma}(\pp,\rr,t)$
defined quantumly through a Wigner transform:
\be
\delta\hat n_\sigma(\pp,\rr,t)=\sum_{\qq}\eee^{\ii\qq\cdot\rr} \delta n_{\sigma}(\pp,\qq,t), \qquad
\delta n_{\sigma}(\pp,\qq,t)\equiv\meanv{\hat a^\dagger_{\pp-\qq/2,\sigma} \hat a_{\pp+\qq/2,\sigma}} -\delta_{\qq\zero} \meanv{\hat a^\dagger_{\pp\sigma} \hat a_{\pp\sigma}}_{\hat\varrho_{\rm th}}
\ee
The external drive couples locally to the density $\rho_\sigma(\rr)=\sum_{\pp} n_\sigma(\pp,\rr)$ through
a spin-dependent field $U_\sigma$:}
\be
\hat H_{\rm d}=\int {\dd^3 r}U_\sigma(\rr,t)\hat\psi_\sigma^\dagger(\rr)\hat\psi_\sigma(\rr)=\sum_{\qq} U_\sigma(\qq,t)\sum_{\pp}\hat a_{\pp+\qq/2,\sigma}^\dagger \hat a_{\pp-\qq/2,\sigma}
\ee
where the Fourier transform of the drive is defined by $U_\sigma(\qq)=\int \dd^3 r \eee^{-\ii\qq\cdot\rr}U_\sigma(\rr)/L^3$.

We explain in Ref.~\cite{heff} how we obtain the time-evolution of $\delta n_{\sigma}(\pp,\qq,t)$ in the Born-Markov approximation (see also Sec. I.4. in \cite{theserepplinger}).
The hierarchy of equations of motion connecting the few-body correlation functions is controlled by the coupling constant $g$, 
such that the one-body correlation functions are of order $g^0$, the two-body correlations of order $g^1$ and so-on. To leading order in $g$, 
the hierarchy can be truncated at the level of one- and two-body correlations,  \textit{i.e.} quadratic   ($\hat a^\dagger \hat a$) 
and quartic ($\hat a^\dagger \hat a^\dagger \hat a \hat a$) operators \cite{Bonitz1998}. 
The separation of scales between the Fermi energy $\epsilon_{\rm F}$ and interaction energy $g \rho$ justifies a Markovian elimination of the quartic operators.
Focusing on long-wavelength physics, we finally retain the leading terms in 
$q/\kF$ in the resulting equation for quadratic operators. 

\begin{figure}
\begin{center}
\begin{tikzpicture}[scale=2]

\def\xmax{4.5}
\def\ymax{4}

\begin{scope}[blend mode=multiply]

\shade[left color=blue!70, right color=blue!0]
    (0,0) rectangle (1.,\ymax);

\shade[bottom color=red!70, top color=red!0]
    (0,0) rectangle (\xmax,1.);


\shade[top color=orange!80, bottom color=orange!0]
    (0,3) rectangle (\xmax,\ymax);

\end{scope}

\filldraw[white]
    (0,0)
    -- plot[domain=0.05:\xmax, samples=200]
       (\x,{1.4*exp(-4.5/\x)})
    -- (\xmax,0)
    -- cycle;
    
\draw[dashed, thick, domain=0.05:\xmax, samples=200]
    plot (\x,{1.4*exp(-4.5/\x)}) node[right] {$T_c/\TF$};

\draw (0.45,2.) node[rotate=90] {\large Weakly-interacting gas};
\draw (2.,0.45) node[rotate=12] {\large Fermi liquid};
\draw (2.3,3.6) node {\large Boltzmann gas};
\draw (4.2,0.2) node {\large Superfluid (BCS side)};
\draw (\xmax/2+0.5,\ymax/2) node {\large BBGKY hierarchy};

\draw[->,thick] (0,0) -- node[below=0.5cm]  {\large $\kF|a|$} (\xmax+0.5,0);
\draw[->,thick] (0,0) -- node[above left=1cm,rotate=90] {\large $T/\TF$} (0,\ymax+0.1) ;
\draw (0,0) node[below left] {$0$};
\draw (0,\ymax) node[left] {$+\infty$};
\draw (\xmax,0) node[below] {$1$};

\end{tikzpicture}
\end{center}
\caption{The regimes in which a spin-$1/2$ Fermi gas with moderate contact interactions obeys a transport equation.
This cartoon applies primarily to the BCS side ($a<0$) of the phase diagram; it can also apply to the BEC side ($a>0$)
provided the atoms remain on the metastable repulsive branch and do not form bosonic dimers.
This article studies the weakly-interacting limit ($\kF a\to0$ at fixed $T/\TF$, blue shade), in which the transport equation is supported
by quantum particles (with a Fermi-Dirac distribution) interacting via the bare interaction. The Fermi liquid
regime ($T/\TF\to0$ at fixed $\kF a$, red shade), where transport is supported by Landau quasiparticles
colliding with an effective interaction \cite{Landau1959}, was studied in Refs.~\cite{devvisco,heff}; it connects with the weakly-interacting gas
in the joint limit $T/\TF\to0$, $\kF a\to0$ (purple shade in the bottom left corner). Then, in the Boltzmann gas
regime ($T/\TF\to+\infty$ at fixed $\kF a$, orange shade), transport is supported by classical particles colliding
with the two-body quantum scattering amplitude (see \textit{e.g.} Refs.~\cite{Smith2005b,Schafer2010}). This regime connects
with the weakly-interacting gas when $T/\TF\to+\infty$ simultaneously with $\kF a\to0$ (upper left corner); it extends to the unitary regime \cite{Nishida2019,Enss2019} ($\kF |a|=+\infty$),
not shown here. Outside those limiting regimes, the transport of the Fermi gas at $T/\TF\approx 1$ and $\kF |a|\gtrsim 1$ is not captured by
a kinetic equation but rather by a BBGKY hierarchy \cite{Bonitz1998,Kira2015,eindhoven}. \label{schema}}
\end{figure}
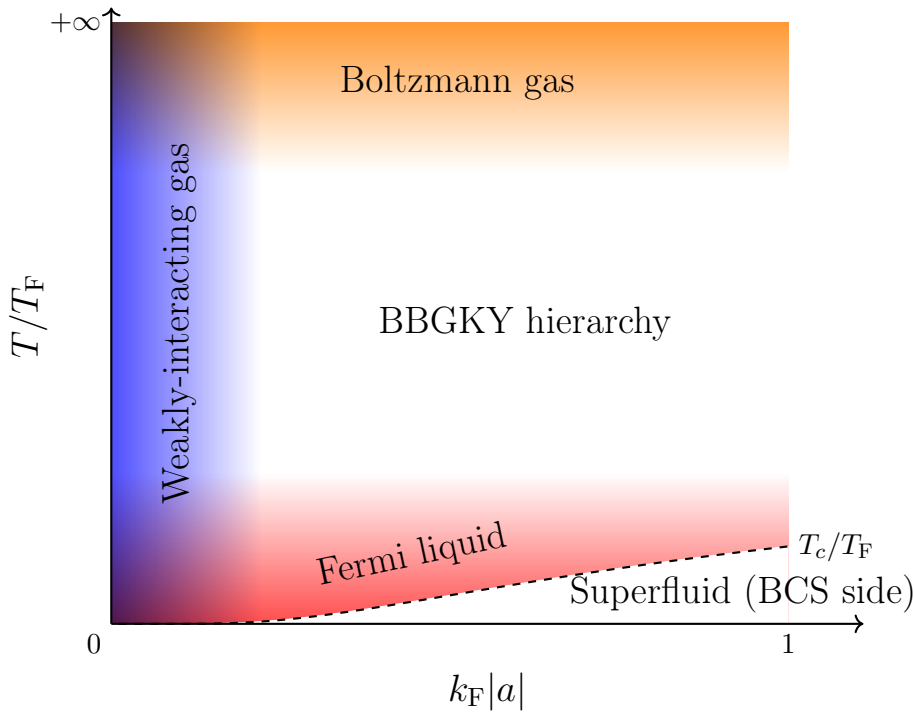

In contrast with the quasiparticle transport discussed in Ref.~\cite{heff}, transport in the weakly-interacting regime is 
supported by the bare particles that experience simply a mean-field shift in their eigenenergy
\be
\epsilon_{\sigma}(\pp)=\frac{p^2}{2m}+g\rho_{-\sigma}^{\rm eq}  \label{epsilonsigma}
\ee
where $\rho_{\upa}^{\rm eq}=\rho_{\dwa}^{\rm eq}=\rho_{{\rm eq}}/2$ are the equilibrium densities, related
to $\mu$ by the equation-of-state (see Section.~\ref{eos}).
This mean-field shift can be absorbed in a redefinition of the chemical potential
\be
\epsilon_{\sigma}(\pp)-\mu=\frac{p^2}{2m}-\mu_0,\qquad \mu=\mu_0+g\rho_{-\sigma}^{\rm eq}
\ee
where $\mu_0$ is the chemical potential of the ideal Fermi gas. \hk{The collision amplitude ($\mathcal{A}$ in the notations of Ref.~\cite{heff})
also coincides with the bare interaction $g$. In Fermi liquids, $\mathcal{A}$
is a low-energy effective interaction among quasiparticles, and follows from the bare interaction by integrating out the high-energy
modes \cite{Senechal1998} or transitions \cite{heff}. In the strongly-interacting virial regime (orange region in Fig.~\ref{schema}),
$\mathcal{A}$ is given by the two-body T-matrix, which describes a renormalization of the bare interaction through multiple
two-body scatterings \cite{Smith2005,Schafer2010,Nishida2019,Enss2019,Hofmann2020}.}

The transport equation is given by:
\be
\ii\frac{\partial{n_{\sigma}}}{\partial t}-\frac{\pp\cdot \qq}{m} \delta n_{\sigma}(\pp,\qq,t)+\frac{\pp\cdot \qq}{m}\frac{\partial n_{\rm eq}}{\partial \epsilon}\Big\vert_{\epsilon=\epsilon_\sigma(\pp)}\bb{g\delta\rho_{-\sigma}(\qq,t)+U_\sigma(\qq,t)}=\ii I_{\rm lin,\sigma}(\pp,\qq,t) \label{transportT}
\ee
where 
\be
\delta\rho_{\sigma}(\qq,t)=\sum_{\pp\in\mathcal{D}} \delta n_{\sigma}(\pp,\qq,t)
\ee
is the fluctuation of the spin-$\sigma$ density.

The collision integral (as in the low temperature case \cite{disphydro,devvisco}) can be decomposed 
into a damping rate $\Gamma$ and two off-diagonal collision kernels\footnote{We have dropped the $\qq$ and $t$ arguments $n_\sigma$ and $I_\sigma$ in \eqqref{Icolllin}.} $E$ and $S$:
\be
I_{\rm lin,\sigma}(\pp)=-\bb{\Gamma(\pp) \delta n_{\sigma}(\pp)+\frac{1}{V}\sum_{\pp'} \bbcro{E(\pp',\pp)n_{-\sigma}(\pp')-S(\pp',\pp) \bb{\delta  n_{\sigma}(\pp')+\delta  n_{-\sigma}(\pp')}}}\label{Icolllin}
\ee
\hk{The various components of the kernel follow from the linearization
of the nonlinear collision integral (\textit{i.e.} for $n_\sigma$ not close to $n_{\rm eq}$):
\begin{multline}
I_{\sigma}(\pp)=
2\pi\frac{g^2}{V} \sum_{\pp_2,\pp_3,\pp_4} \delta_{\pp+\pp_2,\pp_3+\pp_4} 
\delta\bb{\frac{p^2}{2m}+\frac{p_2^2}{2m}-\frac{p_3^2}{2m}-\frac{p_4^2}{2m}} \\
\times \bbcro{n_{\sigma}(\pp) n_{-\sigma}(\pp_2) \bar n_{-\sigma}(\pp_3) \bar n_{\sigma}(\pp_4)-\bar n_{\sigma}(\pp) \bar n_{-\sigma}(\pp_2)  n_{-\sigma}(\pp_3)  n_{\sigma}(\pp_4)} \label{Innlin}
\end{multline}
The linearization of the Fermi-Dirac gain-loss factor (between the square brackets in \eqqref{Innlin}) for $n_\sigma=n_{\rm eq}+\delta n_\sigma$ yields diagonal terms
proportional to $\delta n_{\sigma}(\pp)$ that contribute to $\Gamma(\pp)$, and off-diagonal terms proportional to
$\delta n_{\sigma}(\pp')$, $\pp'\neq\pp$. Among the latter, we distinguish between the ``particle-particle'' terms where
$\pp$ and $\pp'$ are on the same side of the collision ($\pp'=\pp_2$ in \eqqref{Innlin})
that we collect in $E$, and the ``particle-hole'' terms that we collect in $S$ and where $\pp$ and $\pp'$ are on opposite sides  ($\pp'=\pp_3$ or $\pp_4$ in \eqqref{Innlin}):}
\begin{multline}
\Gamma(\pp)=2\pi\frac{g^2}{V^2} \sum_{\pp_2,\pp_3,\pp_4} \delta_{\pp+\pp_2,\pp_3+\pp_4} 
\delta\bb{\frac{p^2}{2m}+\frac{p_2^2}{2m}-\frac{p_3^2}{2m}-\frac{p_4^2}{2m}} \\ \times \bbcro{n_{\rm eq}(\pp_2)(1-n_{\rm eq}(\pp_3)-n_{\rm eq}(\pp_4))+n_{\rm eq}(\pp_3) n_{\rm eq}(\pp_4)} \label{Gammap}
\end{multline}
\begin{multline}
E(\pp,\pp')=2\pi\frac{g^2}{V} \sum_{\pp_3,\pp_4}\delta_{\pp+\pp',\pp_3+\pp_4}\delta\bb{\frac{p^2}{2m}+\frac{p'^2}{2m}-\frac{p_3^2}{2m}-\frac{p_4^2}{2m}} \\ \times \bbcro{n_{\rm eq}(\pp')(1-n_{\rm eq}(\pp_3)-n_{\rm eq}(\pp_4))+n_{\rm eq}(\pp_3) n_{\rm eq}(\pp_4)} \label{Epp}
\end{multline}
\begin{multline}
S(\pp,\pp')=2\pi\frac{g^2}{V} \sum_{\pp_2,\pp_4}\delta_{\pp+\pp_2,\pp'+\pp_4}\delta\bb{\frac{p^2}{2m}+\frac{p_2^2}{2m}-\frac{p'^2}{2m}-\frac{p_4^2}{2m}} \\ \times \bbcro{n_{\rm eq}(\pp_2)(1-n_{\rm eq}(\pp')-n_{\rm eq}(\pp_4))+n_{\rm eq}(\pp') n_{\rm eq}(\pp_4)} \label{Spp}
\end{multline}

We reparametrize the transport equation by factoring out 
the density of available particles ${\partial n_{\rm eq}}/{\partial \epsilon}$ from $n_\sigma$:
\be
\delta n_{\sigma}(\pp,\qq,t)=\frac{\partial n_{\rm eq}}{\partial \epsilon}\Big\vert_{\epsilon_{\rm eq}(\pp)}\nu_\sigma(\pp) 
\label{chgvar}
\ee
In the following, we omit the explicit dependence of $\nu_\sigma$ on $\qq$ and $t$.
The redefined distribution $\nu$ is expected to exhibit a smoother dependence on energy than $n$, so that
this reparametrization is unavoidable in the low temperature regime, where the particle distribution $n_{\sigma}(\pp)$ is peaked about $p_{\rm F}$; elsewhere it is simply convenient for calculation.

In the thermodynamic limit, the transport equation folds onto the integral equation
\begin{multline}
\bb{\omega-\frac{\pp\cdot\qq}{m}} \nu_\sigma({ \pp})+{g}  \frac{\pp\cdot\qq}{m}\int\frac{\dd^3 p'}{(2\pi)^3}  \frac{\partial n_{\rm eq}}{\partial \epsilon}\Big\vert_{\epsilon_{\rm eq}(\pp')} \nu_{-\sigma}({ \pp'})\\
+\ii \bbaco{\Gamma(p)\nu_\sigma(\pp)+\int\frac{\dd^3 p'}{(2\pi)^3}\bbcro{ {E}({\pp},{\pp'})\nu_{-\sigma}({ \pp'})- {S}({\pp},{\pp'})\bbcro{\nu_{\sigma}({ \pp'})+\nu_{-\sigma}({ \pp'})}}}=-\frac{\pp\cdot\qq}{m} U_\sigma
\label{limitethermo}
\end{multline}
where we have assumed a periodic driving $U_\sigma(\qq,t)=U_\sigma(\qq)^{-\ii\omega t}$. In the general case, this equation is
a 3D integral equation; here it simplifies to a 2D equation due to the azimuthal invariance.

\subsection{Dimensionless variables and explicit expression of the collision kernel}
We work in this section in temperature units, that is, with the dimensionless quantities:
\be
\omega_T=\frac{\omega}{v_Tq}, \quad \tilde c=\frac{\omega}{\omega_T},\quad \tilde\mu_0=\frac{\mu_0}{T},\quad \tilde q=\frac{q}{p_{T}},\quad  p=\frac{||\pp||}{p_{T}}, \quad \tilde a={p_{T} a} \label{adim}
\ee
with $v_T=\sqrt{2T/m}$ the typical thermal velocity and $p_T=mv_T$ the associated momentum. Choosing the $z$-axis of the spherical frame along $\qq$, we parametrize the distribution $\nu$ as
\be
\nu_\sigma(\pp)=\nu_\sigma(p,\theta)
\ee
where $p=||\pp||/p_T$ and $\theta=(\widehat{\pp,\qq})$.
\hk{We have exploited the azimuthal invariance of $\nu$, consequence of the driving $U(\qq)$ along
a particular direction.}
Note that we could replace $p$ by an energy
variable $\epsilon=p^2-\tilde\mu_0$. However, since the (anti)symmetry $\epsilon\leftrightarrow-\epsilon$ about the Fermi surface
disappears at nonzero temperature, there is no strong motivation to do so. Instead, keeping $p$ as the radial variable 
will allow us to exploit the differentiability of the distribution $\nu$ in $p=0$.
In dimensionless units, the transport equation takes the form
\begin{multline}
(\tilde{c}- p\cos\theta)\nu_\sigma( p,\theta)-\frac{\tilde a}{\pi^2} p\cos\theta\int \dd^3  p' h( p')\nu_{-\sigma}( p',\theta')+\frac{\ii}{\omega_T \tau}\\\times\bb{\Gamma(p)\nu_\sigma(p,\theta)+\int \dd^3 \tilde p' \bbcro{\tilde{E}( p, p',u)\nu_{-\sigma}( p',\theta')-\tilde{S}( p, p',u)(\nu_\sigma( p',\theta')+\nu_{-\sigma}( p',\theta'))}}=- p\cos\theta  U_\sigma
\end{multline}
where $p'=||\pp'||/p_T$, $h(p)=1/4\text{cosh}^2((p^2-\tilde\mu_0)/2)$, $u=\cos(\widehat{\pp,\pp'})$ and
\be
\tau=\frac{\pi}{2ma^2T^2}
\ee
is the typical collision time in the low-temperature regime. We will generalize
the definition of the mean collision time to arbitrary temperature below (see \eqqref{taum}).

The dimensionless kernels ${E}$ and ${S}$ are computed in Appendix \ref{noyaux}. The result is
\bea
\tilde{E}( p, p',u)&\equiv&\bb{\frac{p_T}{2\pi}}^3\tau E(\pp,\pp')=\frac{1}{ \pi p_+\text{sinh}\frac{\epsilon+\epsilon'}{2}}\frac{\text{ch}\frac{\epsilon}{2}}{\text{ch}\frac{\epsilon'}{2}}\text{ln}\frac{\text{ch}\frac{\epsilon+\epsilon'+  p_+ p_-}{4}}{\text{ch}\frac{\epsilon+\epsilon'-  p_+ p_-}{4}}\label{E}\\
\tilde{S}( p, p',u)&\equiv&\bb{\frac{p_T}{2\pi}}^3\tau S(\pp,\pp')=\frac{1}{2 \pi p_-\text{sinh}\frac{\epsilon-\epsilon'}{2}}\frac{\text{ch}\frac{\epsilon}{2}}{\text{ch}\frac{\epsilon'}{2}}\text{ln}\frac{1+\eee^{p_u^2-\epsilon'}}{1+\eee^{p_u^2-\epsilon}}\label{S}
\eea
with $\epsilon=p^2-\tilde\mu_0$, $\epsilon'=p'^2-\tilde\mu_0$, $p_\pm=\sqrt{p^2\pm2pp'u+p'^2}$, $u_-=\frac{p'-pu}{p_-}$ and $p_u=pp'\sqrt{1-u^2}/p_-$. The expression of $\tilde\Gamma(p)=\tau \Gamma(\pp)$ follows from the conservation law
\be
\tilde\Gamma( p)=-\int \dd^3 p'( \tilde E( p, p',u)-2 \tilde S( p, p',u))\label{Gamma} 
\ee

We introduce the total density and polarisation variables
\be
\nu_\pm=\nu_\uparrow\pm\nu_\downarrow
\ee
This diagonalizes the transport equation with respect to the spin variables:
\begin{multline}
(\tilde{c}- p\cos\theta)\nu_+( p,\theta)-\frac{\tilde a}{\pi^2} p\cos\theta\int \dd^3  p' h(p')\nu_{+}(p',\theta')\\+\frac{\ii}{\omega_T \tau}\bb{\tilde\Gamma(p)\nu_+(p,\theta)+\int \dd^3  p' {(\tilde{E}( p, p',u)-2\tilde{S}( p, p',u))\nu_+( p',\theta')}}=- U_+p\cos\theta \label{nuP}
\end{multline}
\begin{multline}
(\tilde{c}- p\cos\theta)\nu_-( p,\theta)+\frac{\tilde a}{\pi^2} p\cos\theta\int \dd^3  p' h(p')\nu_{-}( p',\theta')\\+\frac{\ii}{\omega_T \tau}\bb{\tilde\Gamma(p)\nu_-(p,\theta)-\int \dd^3  p' {\tilde{E}( p, p',u)\nu_-( p',\theta')}}=- U_- p\cos\theta \label{nuM}
\end{multline}
where $U_\pm=U_\upa\pm U_\dwa$.

\subsection{Projection onto an orthogonal basis}
\label{projection}
We now project those transport equations onto an orthogonal basis of polynomials: 
\be
\nu_\pm(p,\theta)=U_\pm\sum_{l=0}^{+\infty} \nu_{l\pm}(p) P_l(\cos\theta) =U_\pm\sum_{l=0}^{+\infty}\sum_{n=l}^{+\infty} \nu_{n\pm}^l P_l(\cos\theta) Q_n^l(p) \label{expnu}
\ee
Note that $\nu$ is scaled to the intensity $U$ of the drive, such that the coefficients $\nu_{n\pm}^l$
are dimensionless and independent of $U$ in the linear regime $U\to0$.
The smoothness of the distribution $\nu$ 
in $p=0$ implies that $\nu^l$ must be divisible by $p^l$. 
To exploit this property, we construct a different
orthogonal family for each spherical harmonic $l$
through a specific initialization condition:
\begin{align}
&l=0\quad &&Q_0^0=1,\quad &&Q_1^0=p \label{initialization1}\\
&l\geq1 \quad &&Q_{l-1}^l=p^{l-1},\quad &&Q_l^l=p^l \quad (\text{and }Q_{n<l-1}^l=0) \label{initialization2}
\end{align}
The orthogonality condition is however the same for all $l$
\be
\langle Q_n^l,Q_m^l\rangle=\delta_{nm}||Q_n^l||^2
\ee 
with the scalar product weighted by the (dimensionless) density of available states $g$:
\be
\langle  P,Q\rangle=\frac{1}{2}\int_{-\infty}^\infty p^2\dd p\, h(p) P(p) Q(p) \label{prodscalaire}
\ee
The family $\{Q_n^l\}_{n\in\mathbb{N}}$ follows from the usual recurrence relation
\bea
Q_{n+1}^l=p Q_{n}^l-\xi_n^l Q_{n-1}^l
\eea
applied to the initial condition \eqqrefs{initialization1}{initialization2}. 
 We have defined
\be
\xi_n^l=\frac{||Q_n^l||^2}{||Q_{n-1}^l||^2}
\ee

We project the damping rate $ \tilde\Gamma$, and more generally any 1D functions $\tilde F(p)$, according to 
\be
F_{nn'}^l=\frac{1}{2}\int_{-\infty}^\infty p^2\dd p \, h(p) \frac{Q_n^l(p)}{||Q_n^l||^2}\tilde F(p) Q_{n'}^l(p)
\label{projGamma}
\ee
For 3D tensors $\tilde T(\pp,\pp')=\tilde T(p,p',u=(\widehat{\pp,\pp'}))$, such as $\tilde E$ and $\tilde S$, we use
\be
T_{nn'}^l=2\pi\frac{\langle Q_n^l,T^l Q_{n'}^l \rangle}{||Q_n^l||^2} 
\ee
with
\be
T^l(p,p')=\int_{-1}^1\dd u P_l(u) \tilde T(p,p',u) \text{ and } \langle P,T Q \rangle =\frac{1}{2}\int_{-\infty}^{\infty} p^2\dd p\, h(p)p'^2\dd p' {P(p)}T(p,p') Q(p')
\ee
\eqqrefs{nuP}{nuM} projected onto the orthogonal basis become
\begin{multline}
\tilde c\nu_{n\pm}^l-\frac{l}{2l-1}\bb{\nu_{n-1\pm}^{l-1}+\xi_{n+1}^{l-1}\nu_{n+1\pm}^{l-1}}-\frac{l+1}{2l+3}\bb{\nu_{n-1\pm}^{l+1}+\xi_{n+1}^{l+1}\nu_{n+1\pm}^{l+1}}\mp\frac{4\tilde a ||Q_0^0||}{\pi}\nu_{0\pm}^0 \delta_{l1}\delta_{n1}\\+\frac{\ii}{\omega_T\tau}\sum_{n'=0}^{+\infty}\mathcal{M}_{nn'\pm}^l\nu_{n'\pm}^l=-\delta_{l1}\delta_{n1} \label{nunl}
\end{multline}
where the collision term is now set by the matrices
\bea
\mathcal{M}_{nn'+}^l&=&\Gamma_{nn'}^l+E_{nn'}^l-2S_{nn'}^l \label{Mp}\\
\mathcal{M}_{nn'-}^l&=&\Gamma_{nn'}^l-E_{nn'}^l  \label{Mm}
\eea
Numerically, we use the conservation law \eqqref{Gamma} to rewrite $\Gamma(p)$ as an integral similar to the one giving $E_{nn'}^l$ and $S_{nn'}^l$,
and  rewrite the elements of $\mathcal{M}$ in the form
\bea
\mathcal{M}_{nn'\pm}^l&=& 2\pi \int_{-1}^1\dd u \int p^2\dd p\, h(p)p'^2\dd p' \frac{Q_n^l(p)}{||Q_n^l||^2} \tilde T_\pm(p,p',u)\bbcro{P_l(u)Q_{n'}^l(p')-P_0(u)Q_{n'}^l(p)} 
\label{Mnnpm}
\eea
with $\tilde T_+=\tilde E-2 \tilde S$ and $\tilde T_-=-\tilde E$.
The counter-term $P_0(u)Q_{n'}^l(p)$ coming from $\tilde \Gamma$ greatly improves
the numerical accuracy since it compensates the squareroot divergence of
$\tilde E$ and $\tilde S$ in $p=p'$ and $u=\pm 1$.

\subsection{The mean collision time across temperatures}
\begin{figure}
\begin{center}
\includegraphics[width=0.7\textwidth]{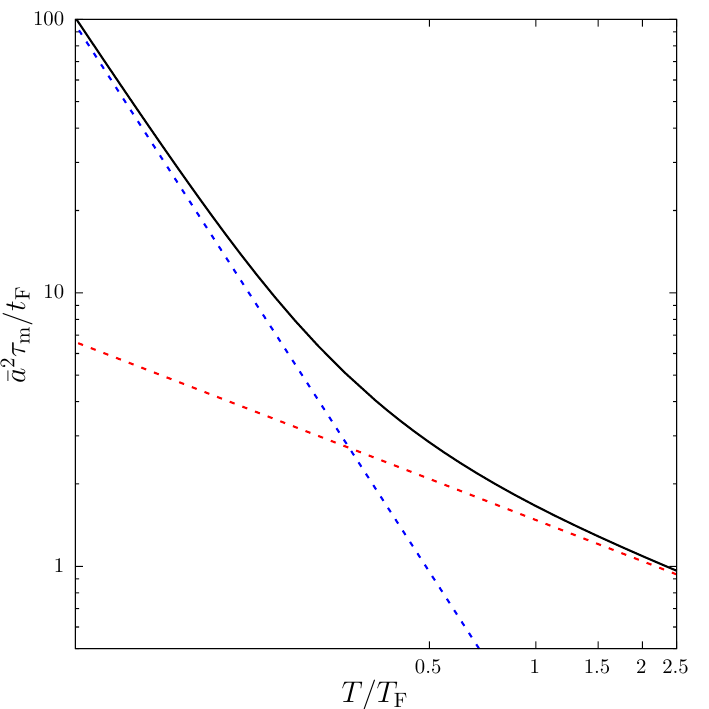}
\end{center}
\caption{The reduced mean collision time $\bar a^2 \tau_{\rm m}/t_{\rm F}$ (with $\bar a=k_{\rm F} a$) 
as function of $T/T_{\rm F}$ at fixed density. It decreases monotonically first as $\tau_{\rm m}\sim (3/4\pi) (T_{\rm F}/T)^2 t_{\rm F}/\bar a^2$
in the low temperature (Fermi liquid) regime (blue dashed curve), then slower, as the collision time $\tau_{\rm m}\sim 1/\sqrt{2} v_{\rm m}\rho_\uparrow \sigma$
of a hard sphere Boltzmann gas of cross section $\sigma=4\pi a^2$, mean velocity $v_{\rm m}=\sqrt{8T/\pi m}$ and only 
$\upa\dwa$ collisions  (corresponding to $\bar a^2 \tau_{\rm m}/t_{\rm F}
\sim (3\pi^2/2)\sqrt{\pi T_{\rm F}/2T}$ in the Fermi units, red dashed curve). Data for Figs.~\ref{figtaum}--\ref{figeta}--\ref{figkappa}--\ref{figD} available in \cite{vtemp}. \label{figtaum}}
\end{figure}

Postponing the solution of the transport equation to Sec.~\ref{hydrointermediaire},
we can already estimate the strength of the collision term ($\propto\mathcal{M}$ in \eqqref{nunl})
relative to the rest of the transport equation. To do so, we introduce the mean collision rate $1/\tau_{\rm m}$ defined as
\be
\frac{1}{\tau_{\rm m}}=\frac{\Gamma_{00}^0}{\tau}=\frac{\int_0^{+\infty}\dd\epsilon \rho(\epsilon) \Gamma(\epsilon)}{\int_0^{+\infty}\dd\epsilon \rho(\epsilon)} \label{taum}
\ee
Mathematically, this definition corresponds to the projection of $\Gamma$ onto
the lowest orthogonal polynomial $Q_0^0$ (see \eqqref{projGamma} for the projection
of the dimensionless rate $\tilde\Gamma=\tau\Gamma$), which we expect to provide the
order of magnitude of the whole collision tensor $\mathcal{M}$. Physically,
$1/\tau_m$ is defined as an energy average of $\Gamma$ (expressed as a function 
of the energy $\epsilon=p^2/2m$)
weighted by the 3D density of available states $\rho(\epsilon)=-\sqrt{\epsilon}\partial n_{\rm eq}/\partial\epsilon$. Plotted in Fig.~\ref{figtaum} as a function of $T/\TF$ at fixed density, the mean collision time
$\tau_m$ is a monotonically decreasing function of $T/\TF$. At low temperature, $\Gamma_{00}^0\underset{T\to0}{\to}{4\pi^2}/{3}$
such that $\tau_{\rm m}$ is comparable to $\tau$ (as expected), albeit with a small numerical prefactor $3/4\pi^2\simeq0.08$. This explains
our use of 
\be
\tau_\sigma=\frac{\tau}{4\pi}=\frac{\pi}{2m\sigma T^2}
\ee
with $\sigma=4\pi a^2$ the scattering cross section.
In practice, $\tau_\sigma$ is numerically closer to $\tau_{\rm m}$ than $\tau$. 

The opposite high-temperature, or
classical limit, is defined by
\be
T\gg \TF \quad \text{ or }\quad z\equiv \eee^{\beta\mu} \to 0
\ee
at fixed scattering length $a$. In this limit, $\Gamma_{00}^0\underset{z\to0}{\sim}2\sqrt{2}z$, such that $\tau_m$ becomes proportional to the collision
time of a classical binary gas of cross-sections $\sigma_{\upa\dwa}=\sigma=4\pi a^2$ and $\sigma_{\upa\upa}=0$,
that is
\be
\tau_{\rm m} \underset{z\to0}{\sim} \frac{\tau_{\rm HT}}{\sqrt{2}} \quad\text{with}\quad \tau_{\rm HT}=\frac{\tau}{2z}=\frac{1}{v_{\rm m} \rho^{\rm eq}_{\upa}\sigma} \label{tauHT}
\ee
We have introduced $v_{\rm m}=\sqrt{8T/\pi m}$, the average velocity of the Boltzmann gas. $\tau_m$ thus interpolates from a $1/T^2$ decay at low $T$ to a $1/\sqrt{T}$
decay at high $T$, as visible in Fig.~\ref{figtaum}.

\section{The hydrodynamic limit}
\label{hydrointermediaire}

\subsection{Equation of state of the weakly-interacting Fermi gas}
\label{eos}
As the equilibrium properties are needed to interpret
the transport properties in the hydrodynamic limit, 
we briefly recall the equation of state of the weakly-interacting Fermi gas.
In particular, we connect the Fermi-Dirac integrals with which this equation of state is usually
expressed to the norm of our polynomials $Q_n^l$.

The equilibrium density and pressure are related to $T$, the coupling constant $g=4\pi a/m$
and $\mu_0=\mu-g\rho_{\rm eq}/2$
 by
\bea
\rho_{\rm eq} &=&  \frac{2}{\lambda_{\rm th}^3} \mathcal{F}_{1/2}=\frac{2}{3\pi^2}p_T^3 W_2 \label{rhoeq}\\
p_{\rm eq} &=& p_{{\rm eq},0}+\frac{1}{4}g\rho_{\rm eq}^2\quad \text{with} \quad p_{{\rm eq},0}= \frac{2T}{\lambda_{\rm th}^3} \mathcal{F}_{3/2} =\frac{4}{15\pi^2} T p_T^3 W_4 \label{peq}
\eea
where $\lambda_{\rm th}=\sqrt{2\pi/mT}$ is the thermal de Broglie wavelength,
and $\mathcal{F}_j=\int_0^\infty t^j\dd t /(\eee^{t-\mu_0/T}+1)/\Gamma(j+1)$ is the complete Fermi-Dirac integral. We have introduced $W_n$,
the norm of $p^{n/2}$ in the scalar product \eqqref{prodscalaire}; it is related to $\mathcal{F}_j$ by:
\be
W_n\equiv||p^{n/2}||^2=\bb{\frac{n+1}{2}}\Gamma\bb{\frac{n+1}{2}}\frac{\mathcal{F}_{(n-1)/2}}{2}
\ee
For the internal energy $u_{\rm eq}$, 
kinetic energy $e_{\rm eq}$, and entropy $s_{\rm eq}$, all per unit volume, we have
\bea
e_{\rm eq} &=&\frac{3}{2} p_{\rm eq,0} \label{eeq}\\
u_{\rm eq}&=& e_{\rm eq} +\frac{1}{4}g\rho_{\rm eq}^2 \label{ueq} \\
s_{\rm eq}&=& \frac{\frac{5}{2}\mathcal{F}_{3/2}-\frac{\mu_0}{T}\mathcal{F}_{1/2}}{\mathcal{F}_{1/2}} \rho_{\rm eq} \label{seq}
\eea
Since the entropy per particle $s_{\rm eq}/\rho_{\rm eq}$ is a function only of ${\mu_0}/{T}$, isentropic transformations
($\dd S=0$) are characterized by $\dd({\mu_0}/{T})=0$. Differentiating \eqqrefs{rhoeq}{peq}, we deduce the isentropic compressibility
$\chi_s=\rho_{\rm eq}(\partial\rho/\partial P)_S$:
\be
\frac{1}{\rho_{\rm eq}\chi_s}=\frac{5T}{3}\frac{\mathcal{F}_{3/2}}{\mathcal{F}_{1/2}}+\frac{g\rho_{\rm eq}}{2}=\frac{2W_4}{3W_2}+\frac{g\rho_{\rm eq}}{2} \label{chiS}
\ee
From \eqqref{eeq}, we can also extract the specific heat
\be
\frac{c_V}{V}\equiv\bb{\frac{\partial e_{\rm eq}}{\partial T}}_{\rho_{\rm eq}}=\frac{3\rho_{\rm eq}}{2}\bb{\frac{W_4}{W_2}-\frac{W_2}{W_0}}=\frac{3\rho_{\rm eq}}{2}\frac{||Q_2^0||^2}{||Q_1^1||^2} \label{cv}
\ee
Finally, the polarisation susceptibility is given by
\be
{\chi_p}\equiv\bb{\frac{\partial (\rho_{\upa}^{\rm eq}-\rho_{\dwa}^{\rm eq})}{\partial (\mu_\upa-\mu_\dwa)}}_{\mu_-=0,T}=\frac{1}{T\lambda_T^3}\mathcal{F}_{-1/2}=\frac{p_T^3}{2T}\frac{W_0}{\pi^2}.
\ee

\subsection{Conserved quantities and collision times}

In hydrodynamic flows, the particle distribution $n$ can be summarized by the conserved
quantities, which are insensitive to collisions, and whose evolution is described by
the Navier-Stokes equations.
There are four conserved quantities, three in the total density sector: $\nu_{0+}^0$, $\nu_{1+}^1$ and $\nu_{2+}^0$ proportional respectively to the total density, the total velocity and the total energy density, and one in the spin sector: $\nu_{0-}^0$, proportional to the polarisation. Mathematically, these conserved quantities are the projection of the distribution $\nu$ onto the four zero eigenfunctions of the collision kernel, namely  $P_1(\cos\theta)Q_1^1(p)=p\cos\theta$, $P_0(\cos\theta)Q_2^0(p)=p^2$, and $P_0(\cos\theta)Q_0^0(p)=1$ which is an eigenfunction in both the total density and spin sector.
The restriction of the transport equation \eqref{nunl} to these conserved quantities gives
\bea
\tilde c \nu_{0+}^0&-&\frac{\xi_1^1}{3}\nu_{1+}^1=0 \label{nu00}\\
\tilde c \nu_{1+}^1&-&\bb{\nu_{0+}^0+\xi_2^0\nu_{2+}^0}-\frac{2}{5}\xi_{2}^2\nu_{2+}^2-\frac{4\tilde a||Q_0^0||^2}{\pi}\nu_{0+}^0=1 \label{nu11} \\
\tilde c \nu_{2+}^0&-&\frac{1}{3}\bb{\nu_{1+}^1+\xi_3^1\nu_{3+}^1}=0 \label{nu02}
\eea
in the total density sector, and
\be
\tilde c \nu_{0-}^0-\frac{\xi_1^1}{3}\nu_{1-}^1 = 0 \label{nu00m}
\ee
in the spin sector.

The non-conserved quantities which introduce dissipation in these systems are $(\nu_{3+}^1,\nu_{2+}^2)$ for the spin-symmetric conserved quantities and $\nu_{1-}^1$ for the polarisation. We shall designate as the \textit{first non-conserved quanties}, those three components of $\nu$, and those that are collisionally coupled to them:
\bea
\vec{\nu}_+^2&=&(\nu_{n+}^2)_{n\geq2} \\
\vec{\nu}_{+,\perp}^1&=&(\nu_{n+}^1)_{n\geq 3}\\
\vec{\nu}_-^1&=&(\nu_{n-}^1)_{n\geq1} 
\eea

In the hydrodynamic limit, the first non-conserved quantities are subleading in $O(\omega_T\tau)$ and can be approximated by
\bea
\vec{\nu}^2_+ &=& -\frac{2\ii}{3}(\omega_T\tau)\nu_{1+}^1 \frac{1}{\mathcal{M}^2_+}\vec{u}_2+O(\omega_T\tau)^2\label{vecnu2}\\
\vec{\nu}^1_{+,\perp} &=& -\ii (\omega_T\tau)\nu_{2+}^0 \frac{1}{\mathcal{M}^{1}_{+,\perp}} \vec{u}_3+O(\omega_T\tau)^2 \label{vecnu1}\\
\vec{\nu}^1_- &=& \ii(\omega_T\tau)\bbcro{1-\nu_{0-}^0\bb{1-\frac{4\tilde a}{\pi}||Q_0||^2}} \frac{1}{\mathcal{M}^1_-} \vec{u}_1+O(\omega_T\tau)^2 \label{vecnu3}
\eea
where $\vec{u}_n=(\delta_{n'n})_{n'\in\mathbb{N}}$ are the unit vectors of our basis. In \eqqref{vecnu1}, $\mathcal{M}^{1}_{+,\perp}=(\mathcal{M}^{1}_+)_{n,n'>3}$ is the restriction of $\mathcal{M}^{1}_+$ to
the subspace orthogonal to $Q_1$ (the zero-eigenfunction associated to the total velocity), where it is then invertible.

From (\ref{vecnu2}--\ref{vecnu3}), we extract the components  $(\nu_{3+}^1,\nu_{2+}^2,\nu_{1,-}^1)$ needed to close the system of equations on the conserved quantities.
Each of them is associated to a distinct projection of $1/\mathcal{M}$, which we write as a collision time:
\bea
\nu_{2,+}^2&=& -\frac{2\ii}{3}\nu_{1+}^1 (\omega_T\tau_\eta) \quad \text{ with } \quad \frac{\tau_\eta}{\tau}=\vec{u}_2 \frac{1}{\mathcal{M}^2_+}\vec{u}_2\label{nu22}\\
\nu_{3,+}^1&=&  -\ii  \nu_{2+}^0 (\omega_T\tau_\kappa) \quad \text{ with } \quad  \frac{\tau_\kappa}{\tau}=  \vec{u}_3 \frac{1}{\mathcal{M}^{1,\perp}_+} \vec{u}_3\label{nu31}\\
\nu_{1,-}^1&=& \ii (1-\nu_{0-}^0) (\omega_T\tau_D)  \quad \text{ with } \quad  \frac{\tau_D}{\tau}=\vec{u}_1\frac{1}{\mathcal{M}^1_-} \vec{u}_1 \label{nu11m}
\eea

\subsection{Relaxation time approximation and higher truncations}
\hk{Below, we shall compare the exact collision times $\tau_\eta$, $\tau_\kappa$, $\tau_D$ to their value $\tau_\eta^0$, $\tau_\kappa^0$, $\tau_D^0$ in the first variational RTA
discussed in the introduction \cite{ChapmanCowling,Smith2005,Schafer2010}. 
Formulated in our vectorial notations, the first approximation is designed to estimate the scalar $\vec{u}\cdot\vec{\nu}$,
where the unknown vector $\vec{\nu}$ solves $\mathcal{M}\vec{\nu}=\vec{u}$. The variational principle is applied
on the 1D space proportional to $\vec{u}$ (that is $\vec{\nu}\approx\lambda\vec{u}$) which yields
\be
\vec{u}\cdot\vec{\nu}\approx\frac{||\vec{u}||^4}{\vec{u}\mathcal{M}\vec{u}}
\ee
For a vector $\vec{u}$ normalized to unity (such as our unit vectors $\vec{u}_n$), this approximates
the collision times in Eqs.~\eqref{nu22}--\eqref{nu11m} through $\vec{u}(\mathcal{M}^{-1})\vec{u}\approx 1/(\vec{u}\mathcal{M}\vec{u})$.}
In our formalism, this approximation is very natural since it consists in truncating the particle distribution to the lowest $Q_n^l$ polynomial:
\be
\nu(p,\theta)\approx\sum_{l=0}^{+\infty} \nu_l^l P_l(\cos\theta) Q_l^l(p) \label{premiereapprox}
\ee
Similarly, the RTA estimates of the collision times are simply the first diagonal coefficients of
the $\mathcal{M}$ matrices:
\bea
 \frac{\tau_\eta^0}{\tau} &=& \frac{1}{\mathcal{M}^2_{22+}}\\
   \frac{\tau_\kappa^0}{\tau} &=&  \frac{1}{\mathcal{M}^{1,\perp}_{33+}}\\
  \frac{\tau_D^0}{\tau} &=& \frac{1}{\mathcal{M}^1_{11-}}
\eea
Systematic improvements of the first approximation are obtained by including
higher $Q_n$ polynomials up to $n_{\rm max}$:
\be
\nu(p,\theta)\approx\sum_{l=0}^{+\infty} \sum_{n=l}^{l+n_{\rm max}} \nu_n^l P_l(\cos\theta) Q_n^l(p) \label{approxnmax}
\ee
The convergence of $\tau_\eta^{n_{\rm max}}$, $\tau_\kappa^{n_{\rm max}}$, $\tau_D^{n_{\rm max}}$
to the exact value is rather slow in the low temperature limit $\tau_\eta^{n_{\rm max}}-\tau_\eta=O(1/n_{\rm max}^2)$
and much faster in the high temperature limit.

\subsection{Navier-Stokes equations at intermediate temperatures}
Reinjecting Eqs.~(\ref{nu22}--\ref{nu11m}) in (\ref{nu00}--\ref{nu02}), we get the Navier-Stokes equations
in projected, dimensionless form:
\bea
\!\!\!\!\!\tilde c \nu_{0+}^0-\frac{\xi_1^1}{3}\nu_{1+}^1\!\!\!&=&\!\!0 \label{nu00hydro}\\
\!\!\!\!\!\tilde c \nu_{1+}^1-\bb{1+\frac{4\tilde a}{\pi}||Q_0^0||^2}\nu_{0+}^0-\xi_2^0\nu_{2+}^0-1 \!\!\!&=&\!\!-\frac{4\ii\omega_T\tau_\eta}{15}\xi_{2}^2 \nu_{1+}^1 +O(\omega_T\tau)^2\label{nu11hydro} \\
\!\!\!\!\!\tilde c \nu_{2+}^0-\frac{1}{3}\nu_{1+}^1\!\!\!&=&\!\!-\frac{\ii\omega_T\tau_\kappa}{3}\xi_3^1 \nu_{2+}^0 \label{nu02hydro} +O(\omega_T\tau)^2\\
\!\!\!\!\!\tilde c \nu_{0-}^0 \!\!\!&=&\!\! \frac{\ii}{3}\xi_1^1\omega_T \tau_D\!-\! \frac{\ii}{3}\xi_1^1\omega_T \tau_D\!\bbcro{1\!-\!\frac{4\tilde a}{\pi}||Q_0||^2}\!\nu_{0-}^0 \label{nu11mhydro} \!+\!O(\omega_T\tau)^2
\eea
where we have gathered the leading terms of ballistic transport on the left-hand side, and the subleading dissipative terms (in $O(\omega_T\tau)$) on
the right-hand side. Since it is hard to recognize the Navier-Stokes equations in this form, we switch back to the physical units by introducing the
usual variables of the hydrodynamical system: the two densities $\rho_\sigma(\rr,t)$, the average velocity ${v}_{\parallel}(\rr,t)$ parallel to $\qq$
and the kinetic energy density $e(\rr,t)$. Since we have worked all along in the linear response regime, our Navier-Stokes equations
are expressed in terms of the linearized fields $\delta\rho_\sigma=\rho_\sigma-\rho_{\rm eq}/2$ and $\delta e=e-e_{\rm eq}$.
The link between the component $\nu_{n,\pm}^l$ of the particle distribution and the hydrodynamic fields is then
\bea
\delta\rho_\pm &\equiv & \int\frac{\dd^3 p}{(2\pi)^3} \bb{n_{\pp\upa}\pm n_{\pp\upa}}=-\frac{||P_0||^2||Q_0^0||^2 p_T^3 U_\pm}{(2\pi)^2 T} \nu_{0,\pm}^0 \label{drho}\\
\rho_0 v_\parallel &\equiv & \int\frac{\dd^3 p}{(2\pi)^3} \frac{\pp \cdot\qq}{m q}\bb{n_{\pp\upa}+n_{\pp\upa}}=-\frac{||P_1||^2||Q_1^1||^2 p_T^2 U_+}{2\pi^2} \nu_{1,+}^1 \label{vparra}\\
\delta e&\equiv& \int\frac{\dd^3 p}{(2\pi)^3} \frac{p^2}{2m} \bb{n_{\pp\upa}+n_{\pp\upa}}=-\frac{||P_0||^2  p_T^3 U_+}{(2\pi)^2} \bb{||Q_2^0||^2\nu_{2,+}^0+||Q_1^1||^2 \nu_{0,+}^0} \label{deltae}
\eea
where $||P_l||^2=2/(2l+1)$.
The last equation on the energy density can be reformulated by eliminating
$\delta e$ in favor of the fluctuation $\delta T$  of the temperature:
\be
\delta e=\bb{\frac{\partial e_{\rm eq}}{\partial T}}_{\rho_+} \delta T+\bb{\frac{\partial e_{\rm eq}}{\partial \rho_+}}_T\delta\rho_+
\ee
Using \eqqrefs{peq}{eeq} and \eqref{cv} from the equation of state, we obtain
\be
{\delta T}=-\frac{U_+}{2}\nu_{2,+}^0
\ee

Converting \eqqrefss{nu11hydro}{nu11mhydro} to the physical variables, and performing the
inverse space-time Fourier transform, we obtain:
\bea
\frac{\partial \rho_+}{\partial t}+\rho_{\rm eq}\frac{\partial v_\parallel}{\partial z}&=&0 \label{NS1}\\
m\rho_{\rm eq}\frac{\partial v_{\parallel}}{\partial t}+\frac{\partial p}{\partial z}-\sum_\sigma \rho_{\sigma}^{\rm eq}\frac{\partial U_\sigma}{\partial z}&=&\frac{4}{3}\eta\frac{\partial^2 v_{\parallel}}{\partial z^2} \label{NS2}\\
\frac{\partial u}{\partial t}+\bb{u_{\rm eq}+p_{\rm eq}}\frac{\partial v_\parallel}{\partial z}&=& {\kappa} \frac{\partial^2 T}{\partial z^2} \label{NS3}\\
\frac{\partial \rho_-}{\partial t} &=& D\frac{\partial^2\rho_-}{\partial z^2} +D{\chi_p}\frac{\partial^2 U_-}{\partial z^2} \label{NS4}
\eea
where $u(\rr,t)=u_{\rm eq}+\delta e(\rr,t)+\frac{1}{2}g\rho_{\rm eq}\delta\rho_+(\rr,t)$ is the local internal energy per unit volume.
We have use the fact the transport of the interaction energy $u-e=g\rho^2/4$ is dissipativeless 
\be
\frac{\partial (u-e)}{\partial t}+(p_{\rm eq}+u_{\rm eq}-p_{{\rm eq},0}-e_{\rm eq})\frac{\partial v_\parallel}{\partial z}=0
\ee
as a consequence of the continuity equation \eqref{NS1}.

A crucial difference with the Navier-Stokes equations at low temperature (see Eqs.~(29--32) in \cite{devvisco})
is the coupling between the temperature and velocity fluctuations, which occurs through the terms ${\partial p}/{\partial z}$
and $\bb{u_{\rm eq}+p_{\rm eq}}{\partial v_\parallel}/{\partial z}$ in \eqqref{NS2} and \eqref{NS3} respectively.
These terms vanish at low temperature since
\bea
p_{\rm eq}=\frac{2}{5}\rho_{\rm eq}\eF+\frac{1}{4}g\rho_{\rm eq}^2 + O(T^2)  \quad \quad \quad \delta p\!\!&=&\!\!\bb{\eF+\frac{1}{2}g\rho_{\rm eq}} \delta \rho + O(T^2) \\
u_{\rm eq}=\frac{3}{5}\rho_{\rm eq}\eF+\frac{1}{4}g\rho_{\rm eq}^2 + O(T^2)  \quad \quad \quad \delta u\!\!&=&\!\!\bb{\eF+\frac{1}{2}g\rho_{\rm eq}} \delta \rho +\frac{c_V}{V}\delta T + O(T^2)
\eea
The momentum balance  \eqqref{NS2} is then decoupled from temperature fluctuations, and the energy balance \eqqref{NS3} folds onto a purely thermal diffusive equation (see Eq.~(31) in \cite{devvisco}). Note that
the absence of terms linear in $T$ in the Sommerfeld expansion of $u_{\rm eq}$ and $p_{\rm eq}$,
can be seen as a consequence of the particle-hole symmetry about the Fermi surface. Without this symmetry,
the thermal and momentum dynamics would remain coupled in the limit $T\to0$.

The three transport coefficients are given by
\begin{align}
\eta&=p_{{\rm eq},0}\tau_\eta &\implies& \qquad \qquad \qquad \bar \eta\equiv \frac{(\kF a)^2}{\rho_{\rm eq}}\eta=\frac{2\pi}{5}\frac{W_4}{W_2}\frac{\TF}{T} \frac{\tau_\eta}{\tau}\\
\kappa &= \frac{||Q_3^1||^2}{||Q_1^1||^2}\frac{\rho_{\rm eq}}{2} \frac{p_T^2}{m^2}\tau_\kappa &\implies& \qquad \qquad \qquad \bar \kappa \equiv \frac{m (\kF a)^2 }{\rho_{\rm eq}}\kappa=\pi  \frac{W_6W_2-W_4^2}{W_2^2} \frac{\TF}{T} \frac{\tau_\kappa}{\tau} \\
D &= \frac{\rho_{\rm eq}}{2m\chi_p} \tau_D &\implies& \qquad \qquad \qquad \bar D\equiv \frac{m(\kF a)^2 }{2} D=\frac{\pi}{3}  \frac{W_2}{W_0}  \frac{\TF}{T} \frac{\tau_D}{\tau}
\end{align}
The first column uses the equation-of-state to provide an expression in terms of physical quantities and collision times (we did not find a simple physical quantity to eliminate the factor $||Q_3^1||^2$ 
in $\kappa$). The second column gives a practical expression of the reduced coefficients $\bar \eta$, $\bar \kappa$ and $\bar D$ which are plotted in Figs.~\ref{figeta}, \ref{figkappa} and \ref{figD}
as a function of 
\be
\theta=\frac{T}{\TF}
\ee 
at fixed density $\rho_{\rm eq}$. 
The coefficients pass through a minimum
in between a $1/\theta^p$ behavior at low $\theta$ ($p=2$ for $\eta$ and $D$, $p=1$ for $\kappa$)
and a $\sqrt{\theta}$ behavior at high $\theta$. The minimum is reached for $\theta_{\rm min}\simeq0.75, 0.3$ and $1$
respectively for $\eta$, $\kappa$ and $D$. Our results for $\eta$ and $\kappa$ are in numerical
agreement with Ref.~\cite{Wu2019} (Fig.~6 therein); the discussion there is however
limited to $z<100$, that is $\theta>0.2$, and thus misses the drastic increase of the deviations $\tau_\eta/\tau_\eta^0-1$ and $\tau_\kappa/\tau_\kappa^0-1$ from the RTA.

\subsection{The Fermi liquid limit}
\label{Fermi}
In the low-temperature limit, the gas behaves as a trivial Fermi liquid, where the Landau quasiparticles
coincide with the bare fermions. The transport coefficients to leading order in $\kF a$
were given explicitly  by Ref.~\cite{disphydro} (see also Ref.~\cite{devvisco} for the spin diffusivity).
The diagonalisation of the collision kernel for an arbitrary collision probability was performed in Refs.~\cite{Sykes1968,Wilkins1968,Pethick1977},
and can be readily applied to the case of an isotropic probability $W\propto g^2$.
Note however that Refs.~\cite{Sykes1968,Wilkins1968} omit the contribution of the quantum force
to dissipation, thus wrongly predicting the viscosity and spin diffusivity \cite{devvisco} to higher orders in $\kF a$.

In the low-temperature regime, we obtain large deviations from the first variational RTA \eqqref{premiereapprox}, up to 25\% for the thermal conductivity:
\begin{alignat}{5}
{\tau_\eta}{} &\simeq& 1.079\tau_\eta^0 +O(\theta),\qquad \qquad \frac{\tau_\eta^0}{\tau} &\underset{T\to0}{\to}& \frac{15}{16\pi^2},\qquad \qquad  \eta&=&0.5153\rho_{\rm eq} \EF\tau_\sigma + O(\theta^{-1}) \label{etaT0}\\
{\tau_\kappa}{} &\simeq& 1.253\tau_\kappa^0 +O(\theta), \qquad \qquad \frac{\tau_\kappa^0}{\tau} &\underset{T\to0}{\to}& \frac{15}{32 \pi^2}, \qquad \qquad  \kappa&=&  0.2493 c_V \vF^2 \tau_\sigma + O(\theta^0) \label{kappaT0}\\
{\tau_D}{}&\simeq&1.124\tau_D^0 +O(\theta), \qquad \qquad \frac{\tau_D^0}{\tau}&\underset{T\to0}{\to}&\frac{9}{16\pi^2}, \qquad \qquad  D &=&0.2682{\vF^2 \tau_\sigma} +O(\theta^{-1}) \label{DT0}
\end{alignat}
where $\theta=T/\TF$ is the reduced temperature and $c_V=m\pF T/3$ is the heat capacity at low temperatures.

\subsection{The Boltzmann gas limit}
\label{Boltzmann}
\begin{figure}
\begin{center}
\includegraphics[width=0.7\textwidth]{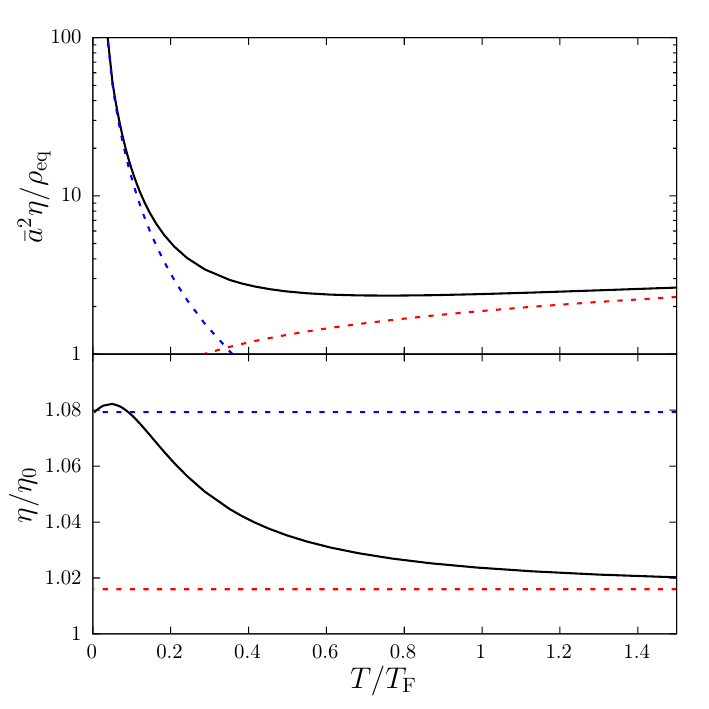}
\end{center}
\caption{(Top panel) The reduced viscosity $\bar\eta\equiv\bar a^2 \eta/\rho_{\rm eq}$ (with $\rho_{\rm eq}=k_{\rm F}^3/3\pi^2$ the total density and $\bar a=\kF a$)
as a function of $\theta=T/T_{\rm F}$ at fixed density. The limiting behaviors are $\bar\eta\underset{\theta\to0}{\sim} (2\pi/5)(\tau_\eta/\tau)/\theta^2\approx 0.1288/\theta^2$
(blue dashed curve) and $\bar\eta\underset{\theta\to+\infty}{\sim} (3\pi^{3/2}/8)(\tau_\eta/\tau_{\rm HT}) \sqrt{\theta}\approx 1.875\sqrt{\theta}$ (red dashed curve).
(Bottom) Ratio of the exact value to the first approximation $\eta/\eta_0=\tau_\eta/\tau_{\eta}^0$ (see \eqqref{premiereapprox}), showing the sharp
deviation from unity at low temperature. \label{figeta}}
\end{figure}

\begin{figure}
\begin{center}
\includegraphics[width=0.7\textwidth]{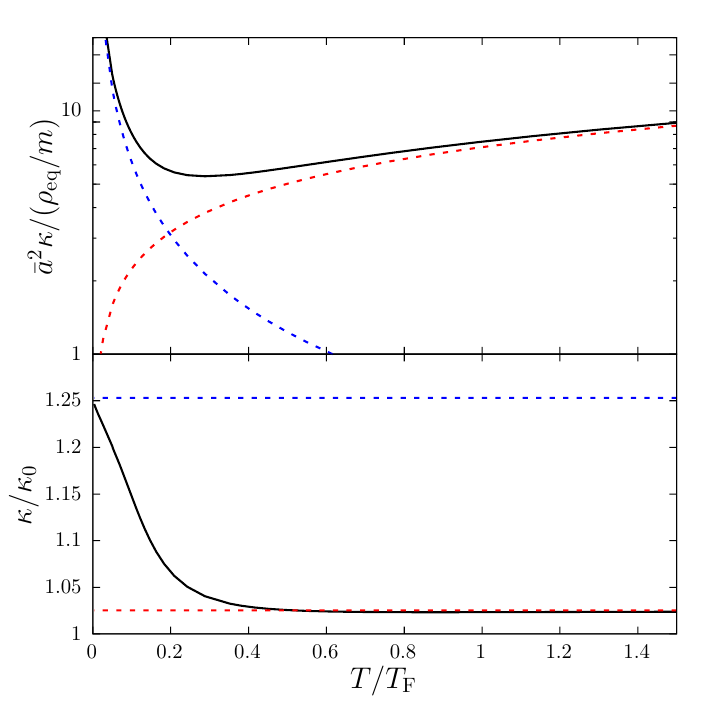}
\end{center}
\caption{(Top panel) The reduced thermal conductivity $\bar\kappa \equiv m\bar a^2 \kappa/\rho_{\rm eq}$
as a function of $\theta=T/T_{\rm F}$ at fixed density. The limiting behaviors are $\bar\kappa\underset{\theta\to0}{\sim} (\pi^3/3)(\tau_\kappa/\tau)/\theta\approx 0.616/\theta$
(blue dashed curve) and $\bar\kappa\underset{\theta\to+\infty}{\sim} (15\pi^{3/2}/16)(\tau_\kappa/\tau_{\rm HT}) \sqrt{\theta}\approx 7.09\sqrt{\theta}$. 
(Bottom) Ratio of the exact value to the first approximation $\kappa/\kappa_0=\tau_\kappa/\tau_{\kappa}^0$ (see \eqqref{premiereapprox}). \label{figkappa}}
\end{figure}

\begin{figure}
\begin{center}
\includegraphics[width=0.7\textwidth]{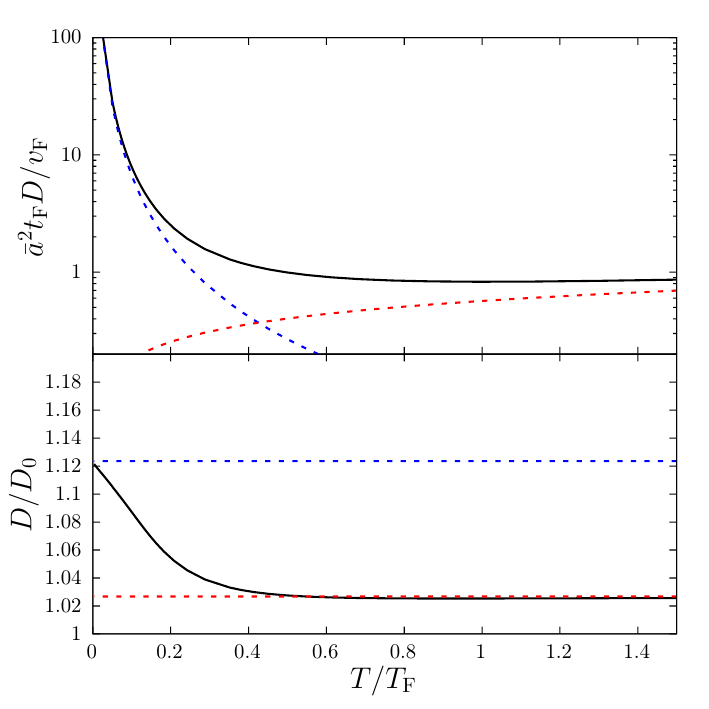}
\end{center}
\caption{(Top panel) The reduced spin diffusivity $\bar D \equiv m\bar a^2 D/2$
as a function of $\theta=T/T_{\rm F}$ at fixed density. The limiting behaviors are $\bar D \underset{\theta\to0}{\sim} (\pi/3)(\tau_D/\tau)/\theta^2\approx 0.067/\theta^2$
(blue dashed curve) and $\bar D \underset{\theta\to+\infty}{\sim} (3\pi^{3/2}/16)(\tau_D/\tau_{\rm HT}) \sqrt{\theta}\approx 0.57\sqrt{\theta}$
(red dashed curve). 
(Bottom) Ratio of the exact value to the first approximation $D/D_0=\tau_D/\tau_{D}^0$ (see \eqqref{premiereapprox}). \label{figD}}
\end{figure}

We now turn to the high-temperature limit $T\gg\TF$. All three collision
times become comparable to $\tau_{\rm HT}$ defined in
\eqqref{tauHT}. We have
\bea
{\tau_\eta} &\simeq& 1.0160{\tau_\eta^0} +O(z)\quad \text{with} \quad \frac{\tau_\eta^0}{\tau_{\rm HT}}\underset{z\to0}{\to}\frac{5}{4\sqrt{2}} \label{tauetaHT}\\
{\tau_\kappa} &\simeq& 1.0253{\tau_\kappa^0} +O(z)\quad \text{with} \quad \frac{\tau_\kappa^0}{\tau_{\rm HT}}\underset{z\to0}{\to}\frac{15}{8\sqrt{2}} \label{taukappaHT}\\
{\tau_D} &\simeq& 1.0268{\tau_D^0}+O(z)\quad \text{with} \quad \frac{\tau_D^0}{\tau_{\rm HT}}\underset{z\to0}{\to}\frac{3\sqrt{2}}{4} \label{tauDHT}
\eea
The exact ratios $\tau_\eta/\tau_\eta^0$ and $\tau_\kappa/\tau_\kappa^0$ agree within $10^{-4}$ with the result of Ref.~\cite{ChapmanCowling} in the ``fourth approximation'' (see Eq.~(10.21, 4) therein). The rapid convergence of $\tau_\eta^{n_{\rm max}}$ and $\tau_\kappa^{n_{\rm max}}$ to  $\tau_\eta$ and $\tau_\kappa$ (see \eqqref{approxnmax}) illustrates the power of the  $Q_n^l$ basis (which folds onto a basis of generalized Laguerre polynomials at high temperatures) in summarizing the energy-dependence of the phase-space distribution. As pointed out by Refs.~\cite{Schafer2010,Smith2005}, the ratios Eqs.~\eqref{tauetaHT}--\eqref{taukappaHT}--\eqref{tauDHT} are
only a few percent away from unity, which contrasts with the low temperature
case. Keeping these exact ratios to express the transport coefficients,
we obtain
\begin{alignat}{4}
\eta&=&\frac{5}{2\sqrt{2}}\frac{\tau_\eta}{\tau_\eta^0}\frac{T}{v_m\sigma},\qquad\qquad \bar \eta&=&\frac{15\pi^{3/2}}{32\sqrt{2}}\frac{\tau_\eta}{\tau_\eta^0}\theta^{1/2}\\
\kappa&=&\frac{75\sqrt{2}}{16}\frac{\tau_\kappa}{\tau_\kappa^0}\frac{T}{m\sigma v_m},\qquad\qquad \bar \kappa&=&\frac{225\pi^{3/2}}{128\sqrt{2}}\frac{\tau_\kappa}{\tau_\kappa^0}\theta^{1/2}\\
D&=&\frac{3\sqrt{\pi}}{8}\frac{\tau_D}{\tau_D^0}\sqrt{\frac{T}{m}}\frac{1}{\rho_{\rm eq}\sigma},\qquad\qquad \bar D&=&\frac{9\pi^{3/2}}{64\sqrt{2}}\frac{\tau_D}{\tau_D^0}\theta^{1/2}
\end{alignat}
which provides the high temperature asymptotic curves (red dashed curves) in Figs.~\ref{figeta}--\ref{figkappa}--\ref{figD}.

\section*{Conclusion}
We introduced a method to exactly solve the quantum Boltzmann equation in the hydrodynamic regime,
based on well-chosen families of orthogonal polynomials. 
We applied this method to obtain exact values of the transport coefficients of a Fermi gas with contact interactions to leading order
in the scattering length $a$, and pointed out large deviations from the relaxation time approximation in the
low-temperature regime.
Owing to its numerical frugality and to its natural truncation scheme, our method is a natural starting point to
explore more complicated transport dynamics, such as second-order hydrodynamics \cite{ChangUhlenbeck,Uribe2008,Schafer2014secondorder,Shukla2024}, collisionless regimes, or nonlinear flows.

\begin{acknowledgements}
H.K. acknowledges support from the French Agence Nationale de la Recherche (ANR), under grant ANR-23-ERCS-0005 (project DYFERCO).
\end{acknowledgements}

\pagebreak
\pagebreak

\appendix
\section{Analytic expression of the collision kernel}
\label{noyaux}
We detail here the derivation of \eqqrefs{E}{S} for the dimensionless collision kernels $\tilde E$ and $\tilde S$:
\begin{multline}
\tilde E(\tilde p,\tilde p',u)=\frac{2}{\pi^2}\int \dd^3\tilde p_3 \dd^3\tilde p_4\delta({\tilde\pp+\tilde\pp'-\tilde\pp_3-\tilde\pp_4})\delta\bb{ {\tilde p^2}+ {\tilde p'^2}- {\tilde p_3^2}- {\tilde p_4^2}} \\ \bbcro{n(\tilde p'^2)\bb{1- n(\tilde p_3^2)-n(\tilde p_4^2))}+n(\tilde p_3^2)n(\tilde p_4^2)}
\end{multline}
\begin{multline}
\tilde S(\tilde p,\tilde p',u)=\frac{2}{\pi^2}\int \dd^3 \tilde p_2 \dd^3 \tilde p_4\delta({\tilde\pp+\tilde\pp_2-\tilde\pp'-\tilde\pp_4})\delta\bb{ {\tilde p^2}+ {\tilde p_2^2}- {\tilde p'^2}- {\tilde p_4^2}} \\
\bbcro{n(\tilde p_2^2)\bb{1- n(\tilde p'^2)-n(\tilde p_4^2))}+n(\tilde p'^2)n(\tilde p_4^2)}
\end{multline}
\paragraph{Calculation of $\tilde E$}
To treat the energy-momentum conservation in $\tilde E$, we introduce
the variables
\be
\tilde\pp_3=\frac{\KK+\yy}{\sqrt{2}} \qquad \tilde\pp_4=\frac{\KK-\yy}{\sqrt{2}} 
\label{changevar}
\ee
Momentum conservation fixes $\KK=(\tilde \pp+\tilde  \pp')/\sqrt{2}$ and energy conservation $y=||\tilde  \pp-\tilde  \pp'||/\sqrt{2}$. 
Locating $\yy$ in a spherical frame of axis $\KK$, there remains to integrate over $u_y=\text{cos}(\KK,\yy)$:
\begin{multline}
\tilde E(\tilde p,\tilde p',u)=\frac{1}{\pi \tilde p_-} \int_{0}^1 \dd u_y \Bigg\{ n(\tilde p'^2)\bbcro{1- n\bb{\frac{\tilde p_+^2+\tilde p_-^2+2\tilde p_+\tilde p_-u_y}{2}}-n\bb{\frac{\tilde p_+^2+\tilde p_-^2-2\tilde p_+\tilde p_-u_y}{2}}}\\{+n\bb{\frac{\tilde p_+^2+\tilde p_-^2+2\tilde p_+ \tilde p_-u_y}{2}}n\bb{\frac{\tilde p_+^2+\tilde p_-^2-2\tilde p_+\tilde p_-u_y}{2}}}\Bigg\}
\end{multline}
where $\pp_\pm=\pp\pm\pp'$. We have used the azimuthal invariance of the integrand, and the symmetry $u_y\leftrightarrow -u_y$. This integral
can be evaluated analytically, which results in \eqqref{E}.

\paragraph{Calculation of $\tilde S$}
For $\tilde S$, where $ \pp$ and $\pp'$ do not play symmetric roles, 
we keep $\pp_2$ as the integration variable and eliminate $\pp_4$ using momentum conservation.
Locating $\pp_2$ in a spherical frame of axis $\pp_-$,
we write the energy conservation constraint in the form:
\be
 p^2+ p_2^2- p'^2-( \pp+ \pp_2- p')^2=2p_- (p_2 u_2 - p'u_-)
\ee
with $u_2=\cos(\pp_2,\pp_-)$ and $u_-=\cos(\pp',\pp_-)=(\tilde p'-\tilde  p u)/\tilde p_-$. 
Using the Dirac delta function to integrate over $u_2$, there remains an integral over $\epsilon_2=p_2^2$ limited to 
$\epsilon_2>p'^2 u_-^2$:
\be
\tilde S(\tilde p,\tilde p',u)=\frac{1}{\pi \tilde p_-}\int_{p'^2 u_-^2}^{+\infty}{\dd \epsilon_2} \Big\{n(\epsilon_2)\bbcro{1- n(p'^2)-n(\epsilon_2+p^2-p'^2)}\\\bbcror{+n(p'^2)n(\epsilon_2+p^2-p'^2)}\Big\}
\ee
which finally results in \eqqref{S}.

\section{Boltzmann gas limit of the projected transport equation}
We detail here the high temperature limit of the transport equation, from which
we obtained the asymptotic behaviors of the transport coefficients in Sec.~\ref{Boltzmann}.

The (dimensionless) collision kernel $\tilde E$ and $\tilde S$, as well as the
damping rate $\tilde \Gamma$, scale as $z=\eee^{\mu/T}$ in this limit:  
\begin{alignat}{4}
\tilde{E}( \tilde p, \tilde p',u)&\underset{z\to0}{\sim}z\tilde{E}_{\rm HT}( \tilde p, \tilde p',u),\qquad\qquad  &\tilde{E}_{\rm HT}( \tilde p, \tilde p',u)&=\frac{\tilde p_-}{\pi}\eee^{-\tilde p'^2}\label{EHT}\\
\tilde{S}( \tilde p, \tilde p',u)&\underset{z\to0}{\sim}z\tilde{S}_{\rm HT}( \tilde p, \tilde p',u),\qquad\qquad &\tilde{S}_{\rm HT}( \tilde p, \tilde p',u)&=\frac{1}{\pi\tilde p_-}\eee^{-\tilde p'^2 u_-^2}\label{SHT}\\
\tilde \Gamma(\tilde p)&\underset{z\to0}{\sim}z\tilde{\Gamma}_{\rm HT}( \tilde p),\qquad\qquad &\tilde{\Gamma}_{\rm HT}( \tilde p)&=\eee^{-\tilde p^2}+\frac{(1+2\tilde p^2)\sqrt{\pi}\text{erf}(\tilde p)}{2\tilde p}\label{GammaHT}
\end{alignat}
where we recall that $\pp_-=\pp-\pp'$ and $u_-=(p'-pu)/p_-$. 
We then rescale the collision time $\tau\to\tau_{\rm HT}=\tau/2z$ (see \eqqref{tauHT}), and redefine the orthogonal
polynomials using a scalar product weighted by the Maxwell-Boltzmann distribution instead of the Fermi-Dirac:
\be
\langle  P,Q\rangle_{\rm HT}=\frac{1}{2}\int_{-\infty}^\infty p^2\dd p \eee^{-p^2} P(p) Q(p) \label{prodscalaireHT}
\ee
The projected transport equation has the same structure as \eqqref{nunl}, except that the mean-field force $\propto g\delta\rho_\sigma$
becomes negligible in this limit:
\begin{multline}
\tilde c\nu_{n\pm}^l-\frac{l}{2l-1}\bb{\nu_{n-1\pm}^{l-1}+\xi_{n+1}^{l-1}\nu_{n+1\pm}^{l-1}}-\frac{l+1}{2l+3}\bb{\nu_{n-1\pm}^{l+1}+\xi_{n+1}^{l+1}\nu_{n+1\pm}^{l+1}}\\+\frac{2\ii}{\omega_T\tau_{\rm HT}}\sum_{n'=0}^{+\infty}\bb{\mathcal{M}_{\rm HT}}_{nn'\pm}^l\nu_{n'\pm}^l=-\delta_{l1}\delta_{n1} \label{nunlHT}
\end{multline}
The high-temperature
version of the collision matrix ${\mathcal{M}_{\rm HT}}$ is expressed as in
\eqqrefs{Mp}{Mm}, with $E\to E_{\rm HT}$, $S\to S_{\rm HT}$ and $\Gamma\to\Gamma_{\rm HT}$.
The decomposition of the functions $\tilde E_{\rm HT}$, $\tilde S_{\rm HT}$ and $\tilde \Gamma_{\rm HT}$
into the corresponding matrices occurs exactly as in Sec.~\ref{projection}, simply replacing the scalar
product by \eqref{prodscalaireHT}.

\providecommand*\hyphen{-}

\end{document}